# Computational aberration correction in spatiotemporal optical coherence (STOC) imaging

Dawid Borycki[+,1], Egidijus Auksorius[+,1], Piotr Węgrzyn[1,2], and Maciej Wojtkowski[*,1]

[1] *Institute of Physical Chemistry, Polish Academy of Sciences, Kasprzaka 44/52, 01-224 Warsaw, Poland*
[2] *Faculty of Physics, University of Warsaw, Pasteura 5, 02-093 Warsaw, Poland*

[+] *Authors contributed equally, \*Corresponding author: mwojtkowski@ichf.edu.pl*

**Spatiotemporal optical coherence (STOC) imaging is a new technique for suppressing coherent crosstalk noise in Fourier-domain full-field optical coherence tomography (FD-FF-OCT). In STOC imaging, the time-varying inhomogeneous phase masks modulate the incident light to alter the interferometric signal. Resulting interference images are then processed as in standard FD-FF-OCT and averaged incoherently or coherently to produce crosstalk-free volumetric OCT images of the sample. Here, we show that coherent averaging is suitable when phase modulation is performed for both interferometer arms simultaneously. We explain the advantages of coherent over incoherent averaging. Specifically, we show that modulated signal, after coherent averaging, preserves lateral phase stability. This enables computational phase correction to compensate for geometrical aberrations. Ultimately, we employ it to correct for aberrations present in the image of the photoreceptor layer of the human retina that reveals otherwise invisible photoreceptor mosaic.**

Optical coherence tomography (OCT) uses interferometric detection to provide high-resolution volumetric imaging of biological samples *in vivo* [1]. Since its original invention in time-domain [2], researchers improved the OCT speed performance by employing the Fourier domain [3] and tunable lasers [4] to reach video-rate acquisition [5]. Though OCT technology exhibited rapid development in recent years, there are still crucial issues. For instance, most scanning OCT systems feature low NA objectives to increase the axial imaging range by extending the depth of field. As a result, the lateral resolution becomes poor compared to axial resolution, making the *in vivo* cellular-level imaging challenging. Additionally, most of OCT systems use a single detection channel and scan the object laterally.

The anisotropic resolution problem was recently tackled by Zhou et al. [6], who extended the commercial OCT system with the angle-dependent tomographic imaging and deep learning-based processing to achieve isotropic resolution of the order of ∼2.5 $\mu$m.

An improved transverse resolution is provided by the full-field (FF) OCT, which uses wide-field illumination and parallel interferometric detection [7]. This approach was originally developed in time-domain with spatially incoherent light sources (LED or thermal sources). Such FF-OCT was shown to provide nearly isotropic resolution below 1 $\mu$m [8]. However, an attempt to further boost the FF-OCT imaging speed with Fourier-domain detection that uses tunable lasers resulted in a crosstalk problem. Namely, the spatial coherence of the laser produces coherent artifacts – the so-called crosstalk-generated noise, which induces the speckle pattern and hampers the image resolution [9]. Though Fourier-domain (FD-) FF-OCT, supported by numerical phase correction, reveals cellular features of the retina, crosstalk prevents visualization of the deeper retinal layers like choroid [10].

Recently, we developed the novel approach to suppress the coherent crosstalk in FD-FF-OCT, which we call spatiotemporal optical coherence (STOC) manipulation [11, 12]. In STOC, we modulate the phase of light incident on the sample in time with a set of phase patterns generated by the spatial phase modulator (SPM). Then, the phase-modulated 3D reconstructions are averaged incoherently to achieve the crosstalk-free volumes. Recently, we used a liquid crystal spatial light modulator (LC SLM) as the SPM to image the 1951 USAF resolution test chart covered by biological tissue *ex vivo* [13]. In parallel, we developed a crosstalk-free FD-FF-OCT system that employed a high-speed deformable membrane as an SPM [14]. We already applied it to image human skin [14] and retina [15] *in vivo*.

In this contribution, we show that crosstalk free FD-FF-OCT can be described with the same formalism that we previously used for STOC with the only difference that we use the coherent instead of incoherent averaging [11]. Moreover, we show that such an approach makes it possible to explain the behavior of a stable phase within imaging depth, which in turn allows using the algorithms for digital correction of optical aberrations [16].

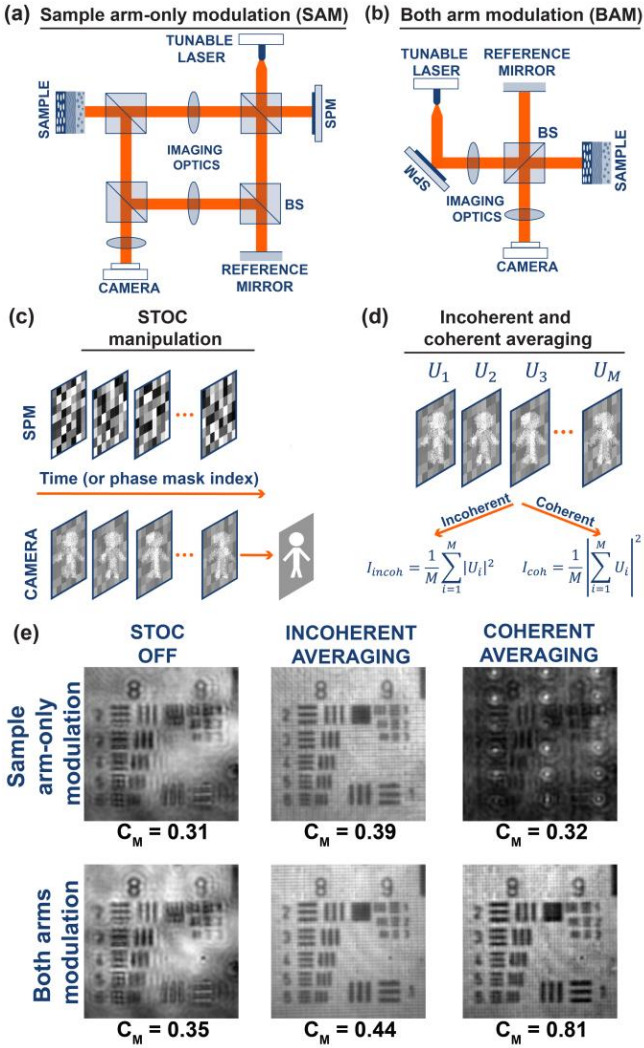

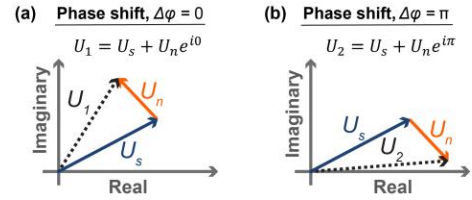

Fig. 2. Two STOC-manipulated phasors with relative phase shifts of $\Delta\varphi = 0$ (a) and $\Delta\varphi = \pi$ (b).

Fig. 1. Two implementations of the STOC imaging in the form of FD-FF-OCT that utilizes SPM. (a) Mach-Zehnder interferometer (MZI) with the SPM placed in the sample arm. (b) Linnik interferometer (LI) with the SPM located before interferometer. (c) In STOC, we modulate the FD-FF-OCT signal in time with a set of $M$ (largely) uncorrelated phase patterns. (d) We can average the resulting signal ($U_1, U_2, ... U_M$) incoherently or coherently. However, only in the LI arrangement, the phase relation between the two arms is preserved, and we can benefit from the coherent averaging to improve the contrast, $C_M$ of the final image (e). Abbreviations: SPM – spatial phase modulator; BS – beam splitter, $C_M$ – Michelson's contrast.

Figure 1(a,b) sketches the two optical setups for STOC imaging [13, 14]. The setups are based on the FD-FF-OCT configuration that uses SPM for phase modulation. The SPM can be placed in either the sample arm [Fig. 1(a)], or before the interferometer [Fig. 1(b)]. We call the former sample arm-only modulation (SAM), and the latter both arm modulation (BAM). In both cases, the SPM sequentially displays uncorrelated phase patterns, each of which is active during the whole laser sweep [Fig. 1(c)]. We record such modulated interferometric images with a two-dimensional camera and apply standard OCT data processing.

Specifically, we average the resulting 3D reconstructions on either the magnitude- (incoherent averaging) or amplitude-basis (coherent averaging) [Fig. 1(d)]. However, when we use SAM [Fig. 1(a)], we distort the phase relation between the reference and sample fields. Thus, in this mode, we can only use incoherent averaging. In the second case, when we use BAM modulation, the phase relation between the two arms is preserved since the signal is both arms is altered by the same phase patterns. Consequently, as in the dynamic phase microscopy, the interference occurs only between the light coming from the sample layer, for which focusing of the back-reflected speckle pattern is the same as for the reference arm [17]. Hence, the coherent averaging is possible and, as explained below, we achieve improved image contrast, compared to the incoherent averaging.

We validated the above approach experimentally, as shown in Fig. 1(e). To this end, we used LC SLM as SPM and imaged the 1951 USAF resolution test chart covered with a tailored micro diffuser (with 1° angle). When there is no phase modulation – STOC is disabled (STOC OFF column) – we get a distorted sample image due to the photon paths deflected by the diffuser. Then, we employed STOC manipulation to modulate phase with $M = 128$ patterns derived from the Hadamard matrix. For both hardware configurations, shown in Fig. 1(a,b), we suppress image distortions by STOC manipulation with the incoherent averaging [second column in Fig. 1(e)]. However, corresponding images contain an additive offset from an incomplete rejection of the crosstalk noise. To overcome this issue, we use BAM and coherent averaging [third column in Fig. 1(e)]. On the other hand, for coherent averaging in SAM, we do not see the sample image but only residuals and regular grid from the Hadamard phase patterns.

To explain the difference between the incoherent and coherent averaging, we use phasor analysis. Figure 2 depicts two phasors $U_1, U_2$ of the STOC-manipulated signal with relative phase shifts of $\Delta\varphi = 0$ and $\Delta\varphi = \pi$, respectively:

$$U_1 = U_s + U_n e^{i0}, \quad U_2 = U_s + U_n e^{i\pi}, \qquad (1)$$

where, $U_s$ is the useful signal and $U_n$ is the crosstalk noise, which represents the effect of the diffuser. We assume that only the second term ($U_n$) changes upon the modulation because by definition, $U_s$ is related to image-bearing photons that propagate the most direct paths. These light paths are not altered by the phase modulation. Under this assumption we average $U_1$ and $U_2$, defined in Eq. (1), incoherently:

$$I_{incoh} = \tfrac{1}{2}\{|U_1|^2 + |U_2|^2\}$$
$$= \tfrac{1}{2}\{2|U_s|^2 + 2|U_n|^2 + 2Re[U_s^* U_n(e^{i0} + e^{i\pi})]\}$$
$$= \tfrac{1}{2}\{2|U_s|^2 + 2|U_n|^2\} = I_s + I_n,$$

and coherently:

$$I_{coh} = \frac{1}{2}|U_1 + U_2|^2$$

$$= \frac{1}{2}|U_s + U_n e^{i0} + U_s + U_n e^{i\pi}|^2$$
$$= \frac{1}{2}|2U_s + U_n(e^{i0} + e^{i\pi})|^2 = 2I_s.$$

In the above derivations, we used an identity $z + z^* = 2\text{Re}[z]$, and definitions $I_{s,n} = |U_{s,n}|^2$.

We can now see that in both cases the cross-terms (containing a product $U_s^* U_n$) cancel out. Hence, we suppress the coherent image distortions induced by the diffuser. However, the additive noise, $I_n$ is not removed in case of incoherent averaging due to an incomplete rejection of the noise intensity ($I_n$). As shown previously in Fig. 1(e), this offset reduces the image contrast. For that reason, the resolution chart features appear gray instead of black. Coherent averaging ($I_{coh}$) overcomes the unwanted offset. Namely, we enhance the signal by a factor of 2, and $I_n$ is not present. As a result, the USAF features appear black as expected [bottom right cell in Fig. 1(e)]. Such a reduction of the noise floor is consistent with the complex averaging in scanning OCT systems [18].

To quantify image contrast improvement, we apply the Michelson's formula, $C_M = (I_{max} - I_{min})/(I_{max} + I_{min})$ to pixels at the edge of the black square, located between eighth and ninth USAF groups. The contrast values, depicted in Fig. 1(e) shows that coherent averaging as implemented in BAM enhances the contrast nearly 2.3 times with respect to the unmodulated case. On the contrary, the incoherent averaging provides only 1.26 times better contrast for both SAM and BAM.

In practice, we use $M$ two-dimensional phase modulation patterns, $\boldsymbol{\varphi}_m = \varphi_m(x, y)$, and images are detected by a two-dimensional array of pixels [$\boldsymbol{U}_{s,n} = U_{s,n}(x, y)$]. Assuming the one-to-one correspondence between SPM and detector pixels, we extend the above derivations to matrix averaging. We denote matrices with bolded symbols and $x, y$ are discrete indices running from 1 to $N$, where $N$ is the detector side length (we assume square detector). After averaging $M$ phasors $\boldsymbol{U}_m = \boldsymbol{U}_s + \boldsymbol{U}_n e^{i\varphi_m}$ with an additive noise term, we obtain:

$$\boldsymbol{I}_{incoh}^{(M)} = \frac{1}{M}\sum_{m=1}^{M}|\boldsymbol{U}_m|^2 = \frac{1}{M}\sum_{m=1}^{M}|\boldsymbol{U}_s + \boldsymbol{U}_n e^{i\varphi_m}|^2$$
$$= \frac{1}{M}\sum_{m=1}^{M}\{|\boldsymbol{U}_s|^2 + |\boldsymbol{U}_n|^2 + 2\text{Re}[\boldsymbol{U}_s^*\boldsymbol{U}_n e^{i\varphi_m}]\}$$
$$= \boldsymbol{I}_s + \boldsymbol{I}_n + \frac{2}{M}\text{Re}\left[\boldsymbol{U}_s^*\boldsymbol{U}_n \sum_{m=1}^{M} e^{i\varphi_m}\right].$$

We perform similar calculations for the coherent averaging:

$$\boldsymbol{I}_{coh}^{(M)} = \frac{1}{M}\left|\sum_{m=1}^{M}\boldsymbol{U}_m\right|^2 = \frac{1}{M}\left|\sum_{m=1}^{M}[\boldsymbol{U}_s + \boldsymbol{U}_n e^{i\varphi_m}]\right|^2$$
$$= \frac{1}{M}\left[\sum_{m=1}^{M}\boldsymbol{U}_s + \boldsymbol{U}_n \sum_{m=1}^{M} e^{i\varphi_m}\right]^2$$

If we now adjust the phase patterns such that $\sum_{m=1}^{M} e^{i\varphi_m} = 0$ we force the noise component to rotate in the complex plane such that it vanishes after averaging. Thus, we obtain $\boldsymbol{I}_{incoh}^{(M)} = \boldsymbol{I}_s + \boldsymbol{I}_n$ and $\boldsymbol{I}_{coh}^{(M)} = M\boldsymbol{I}_s$.

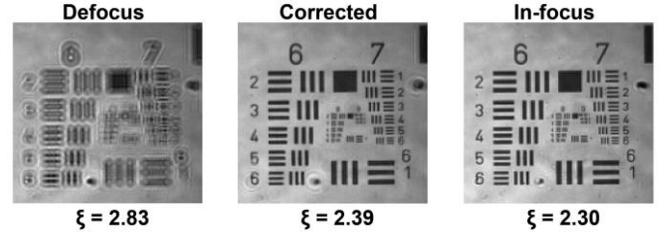

Fig. 3. Numerical phase correction compensates for the defocus aberration in STOC imaging to achieve nearly the same resolution as in-focus. All images were acquired with FD-FF-OCT and STOC manipulation.

Coherent averaging thus enhances the signal term by a factor of $M$ (the number of phase patterns), which agrees with previous studies [19, 20]. In contrast, the incoherent averaging does not enhance the signal and contains an additive offset ($\boldsymbol{I}_n$). However, in both cases, the crosstalk noise is suppressed.

Most importantly, the coherent averaging does not alter the phase of the signal component ($U_s$), if the lateral stability is preserved during the measurement. We can utilize that to correct for geometrical aberrations in the post-processing [10]. We show this here on images acquired with the previously described system [14] that employed a rapid deformable membrane as the SPM to generate pseudo-random phase patterns. The camera implicitly performed coherent averaging since many uncorrelated interference patterns were displayed by the membrane during an acquisition time (16 µs) of a single frame. To induce geometrical aberrations the objective lens was shifted from the optimal focus position by 150 $\mu m$ producing a distorted OCT image, shown in the left column of Fig. 3.

To correct for the defocus, we use digital aberration correction (DAC), which proceeds as follows. The complex data of each sample layer, $U_l(x, y)$ is 2D-Fourier transformed to obtain the spatial spectrum $\widetilde{U}_l(k_x, k_y)$. Then, $\widetilde{U}_l(k_x, k_y)$ is multiplied by $\exp[i\alpha Z_2^0(k_x, k_y)]$, where $\alpha$ is an adjustable parameter, and $Z_2^0$ denotes the Zernike polynomial corresponding to defocus (with OSA/ANSI index of 4). The resulting product $\widetilde{U}_l(k_x, k_y)\exp[i\alpha Z_2^0(k_x, k_y)]$ is 2D-inverse-Fourier transformed to obtain phase-corrected data $U_{l,corr}(x, y)$. We then calculate the image sharpness metric $\xi(\alpha)$ on $|U_{l,corr}(x, y)|^2$. This process is continued for various $\alpha$ until we optimize $\xi(\alpha)$. The resulting, optimized USAF image appears in the second column of Fig. 3. Digital correction leads to an almost diffraction-limited in-focus reference image, depicted in the last column of Fig. 3.

When imaging simple objects like the USAF resolution test chart, the noise offset can be subtracted, and the image contrast can be stretched. However, this becomes an issue for complex objects, especially when imaging through the scattering layer. To show this, we imaged lens tissue with and without STOC manipulation, shown in Fig. 4. Again, the images were acquired with induced defocus. When STOC is disabled the sample OCT image is significantly corrupted by the crosstalk noise. The features of the sample like fibers are barely seen because of the strong influence from noise term $I_n$. However, we can significantly improve OCT image contrast by enabling STOC, which in practice is carried out by activating the deformable membrane or any other fast SPM in the FD-FF-OCT system. Nevertheless, fibers are still blurred due to the defocus aberration.

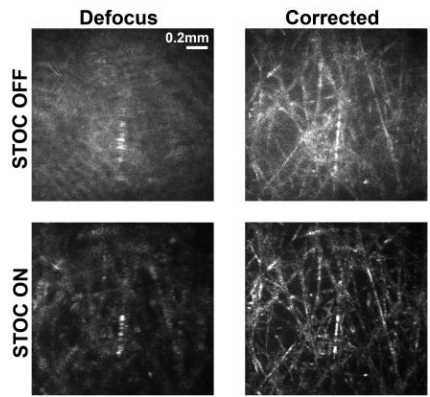

Fig. 4. Removal of crosstalk noise and optical aberrations in images of lens tissue. Crosstalk is removed by STOC (bottom row), whereas defocus aberrations are removed computationally (right column).

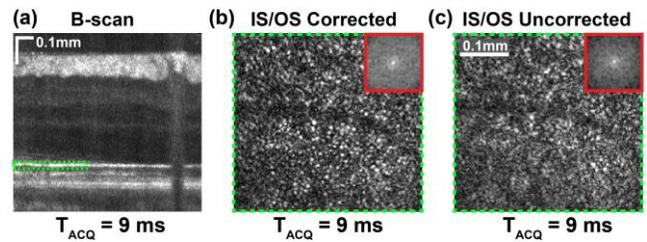

Fig. 5. STOC imaging of the photoreceptor layer of the human retina *in vivo* (Visualization 1). (a) The cross-section (B-scan) with a green dashed rectangle indicating the IS/OS layer, which was corrected computationally to reveal photoreceptor cones (b) that were otherwise invisible (c). Insets in (b,c): 2D power spectrums of the corresponding *en face* images.

Notably, we can correct for that using our DAC algorithm. As shown in the second column of Fig. 4, this algorithm works well on the noisy data (STOC OFF row). The blurred fibers become sharp, and we can see features that were previously invisible. Most importantly, the same approach can be applied to STOC manipulated signal to improve not only the image contrast but also the sharpness of the sample features. This proves that STOC manipulation with coherent averaging implemented by using a fast SPM before the interferometer preserves the signal phase.

Finally, we imaged retina of the healthy 44-years-old volunteer *in vivo* with the FD-FF-OCT system described previously [15]. However, we used only the single volume that was captured within ~9 ms. As shown in Fig. 5(a) we can image deeper into retina than previous FD-FF-OCT systems [10] due to crosstalk suppression. Then, in post-processing, we applied the DAC algorithm, extended to higher-order Zernike polynomials. By doing so we reveal photoreceptor mosaic [Fig. 5(b)], which is not visible without DAC [Fig. 5(c)] nor averaging several volumes [15]. We confirm that observed cellular structure comprises photoreceptor cones using the power spectrum of the *en face* images. The insets of Fig. 5(b,c) show that DAC allows us to see Yellot's ring, whose radius is inversely proportional to the cone spacing.

In summary, by extending FD-FF-OCT with a spatiotemporal phase modulator (SPM), we could average the resulting data incoherently or coherently. We applied this capability to the two hardware implementations, in which the SPM was placed either in the sample arm or before the interferometer. We demonstrated that coherent averaging is suitable in the latter configuration because the phase relation between the two interferometer arms is preserved. We have shown that the phase of the useful signal is maintained after the modulation and the coherent averaging allowing correction of phase errors in post-processing. We employed this to depict the IS/OS layer of the human retina *in vivo*, revealing the photoreceptor mosaic, the primary sensing element of the human visual system. Our results can thus pave the way for FD-FF-OCT to *in vivo* cellular-level non-invasive volumetric imaging.

**Funding.** National Science Center (NCN, 2016/22/A/ST2/00313); European Union's Horizon 2020: research and innovation program (666295);

**Acknowledgments.** We thank Michał Hamkało for his help in the initial stages of this project.